\documentclass[twocolumn,10pt]{IEEEtran}

\usepackage{amsmath,amssymb,amsthm}
\usepackage{graphicx}
\usepackage{xcolor}
\usepackage[acronym]{glossaries}

\newacronym{bd-ris}{BD-RIS}{beyond-diagonal RIS}
\newacronym{bd-fis}{BD-FIS}{beyond-diagonal FIS}
\newacronym{d-ris}{D-RIS}{diagonal RIS}
\newacronym{d-fis}{D-FIS}{diagonal FIS}
\newacronym{em}{EM}{electromagnetic}
\newacronym{fis}{FIS}{fixed intelligent surface}
\newacronym{iid}{i.i.d.}{independent and identically distributed}
\newacronym{los}{LoS}{line-of-sight}
\newacronym{mimo}{MIMO}{multiple-input multiple-output}
\newacronym{ris}{RIS}{reconfigurable intelligent surface}
\newacronym{siso}{SISO}{single-input single-output}
\newacronym{sre}{SRE}{smart radio environment}
\newacronym{ula}{ULA}{uniform linear array}
\newacronym{xpd}{XPD}{cross-polar discrimination}

\begin{document}

\title{\LARGE Movable Signals with Dual-Polarized Fixed Intelligent Surfaces:\\Beyond Diagonal Reflection Matrices}

\author{Matteo~Nerini,~\IEEEmembership{Member,~IEEE},
        Bruno~Clerckx,~\IEEEmembership{Fellow,~IEEE}

\thanks{This work has been supported in part by UKRI under Grant EP/Y004086/1, EP/X040569/1, EP/Y037197/1, EP/X04047X/1, EP/Y037243/1.}
\thanks{Matteo Nerini and Bruno Clerckx are with the Department of Electrical and Electronic Engineering, Imperial College London, SW7 2AZ London, U.K. (e-mail: m.nerini20@imperial.ac.uk; b.clerckx@imperial.ac.uk).}}

\maketitle

\begin{abstract}
This paper investigates wireless systems aided by dual-polarized intelligent surfaces.
We compare \gls{ris}, which adjust their reflection matrices, with movable signals operating with \gls{fis}, which adjust the signal frequency while the surface properties remain fixed.
For both \gls{ris} and \gls{fis}, we consider surfaces with a diagonal reflection matrix, named diagonal \gls{ris}/\gls{fis}, and surfaces with a reflection matrix not limited to being diagonal, named beyond-diagonal \gls{ris}/\gls{fis}.
Movable signals with \gls{fis} always outperform \gls{ris}, achieving at least a fourfold gain.
When transmitter and receiver polarizations differ, beyond-diagonal \gls{fis} further enhances performance.
\end{abstract}

\glsresetall


\section{Introduction}

\Glspl{ris} have been thoroughly investigated as a promising means to enable \glspl{sre} by shaping and controlling the wireless channel \cite{wu21,dir19}.
A \gls{ris} is a surface composed of reflecting elements with adjustable \gls{em} properties, thereby steering incident signals toward desired directions.
Although RIS can offer interesting performance gains, its practical deployment remains challenging.
Dynamic reconfiguration of the \gls{ris} requires dedicated control circuitry with non-negligible power consumption.
An effective \gls{ris} reconfiguration relies on accurate channel knowledge, demanding additional training overhead.
Moreover, since a RIS alters all incident \gls{em} signals, the resulting unpredictable interference from different operators raises concerns.

To circumvent these limitations, an alternative approach has been introduced: movable signals operating with \gls{fis} \cite{ner25b}.
Instead of reconfiguring the \gls{em} properties of the surface, this approach reconfigures the carrier frequency of the transmitted signal, while using a surface of uniformly spaced reflecting elements with fixed \gls{em} properties.
Prior work has demonstrated that, in \gls{siso} systems, movable signals with \gls{fis} can enhance the channel strength by up to four times compared with \gls{ris} \cite{ner25b}.
Although this gain relies on a suitable spectrum availability (ideally, the available carrier frequencies should be distributed over a wide range), it remarkably does not require any reconfigurable hardware or high channel estimation overhead.

Previous work investigated \gls{fis} with a diagonal reflection matrix, which we denote as \gls{d-fis}, mirroring the assumption adopted for conventional \gls{ris}, also known as \gls{d-ris} \cite{ner25b,li25}.
Nevertheless, recent developments in \gls{bd-ris} highlight the benefits of artificially coupling the \gls{ris} elements to each other, which leads to more flexible reflection capabilities.
Extending this concept to \gls{fis} and exploring the performance benefits of \gls{bd-fis}, i.e., \glspl{fis} with fixed reflection matrices not limited to be diagonal, remains unexplored.

In this work, we investigate a practical version of \gls{fis} where the antenna elements are dual-polarized, and show that \gls{bd-fis} can be particularly beneficial in this case.
Dual-polarization has been introduced in \gls{mimo} communications to allow more antennas within a limited space, and increase diversity \cite{oes08}, and it is reasonable to expect that it will also be considered for intelligent surfaces.
Focusing on \gls{siso} systems aided by dual-polarized intelligent surfaces, we analyze and compare the performance of four designs: \gls{d-ris}, \gls{bd-ris}, \gls{d-fis}, and \gls{bd-fis}, where \glspl{fis} are used jointly with movable signals.

Our contributions are the following.
\textit{First}, we characterize the received power achievable with these four designs in the two cases where the transmitter and receiver have the same and opposite polarization.
In this analysis, we consider the effect of the specular reflection due to the structural scattering of the surface, which is often neglected in the \gls{ris} literature, but plays an important role \cite{nos24,abr24,ner24b}.
\textit{Second}, we obtain in closed form the performance gain of \gls{bd-ris}, \gls{d-fis}, and \gls{bd-fis} over \gls{d-ris}.
Movable signals with \gls{fis} always outperform \gls{ris}, achieving at least a four times higher received power.
Furthermore, when the transmitter and receiver have opposite polarization, \gls{bd-fis} yields even larger performance improvements, reaching infinite gains as the inverse of the \gls{xpd} approaches zero.

\section{Dual-Polarized RIS/FIS-Aided Channel Model and Problem Formulation}

Consider a \gls{siso} communication system, as illustrated in Fig.~\ref{fig:system}.
To enhance the channel strength, a reflecting surface made of $N$ dual-polarized elements is deployed, which can be either a \gls{ris} or a \gls{fis}, depending on whether its reflecting properties are reconfigurable or fixed.
Letting the transmitted signal be $x\in\mathbb{C}$, the received signal writes as $y=hx+n$, where $h\in\mathbb{C}$ is the wireless channel and $n\in\mathbb{C}$ is the noise.

\subsection{Channel Model}

Following previous work \cite{wu21,ner25b}, the channel $h$ is given by $h=\mathbf{h}_{R}\boldsymbol{\Theta}\mathbf{h}_{T}-\mathbf{h}_{R}\mathbf{h}_{T}$, where $\boldsymbol{\Theta}\in\mathbb{C}^{N\times N}$ is the reflection matrix of the $N$-element surface, $\mathbf{h}_{R}\in\mathbb{C}^{1\times N}$ is the channel between the surface and receiver, and $\mathbf{h}_{T}\in\mathbb{C}^{N\times 1}$ is the channel between the transmitter and surface.
In $h$, the term $\mathbf{h}_{R}\boldsymbol{\Theta}\mathbf{h}_{T}$ accounts for the signal reflected by the surface and controlled by $\boldsymbol{\Theta}$, while the term $-\mathbf{h}_{R}\mathbf{h}_{T}$ accounts for the signal specularly reflected by the surface (an effect of the structural scattering which is independent of $\boldsymbol{\Theta}$ \cite{nos24,abr24,ner24b}).
The direct link between the transmitter and receiver is considered to be very weak and hence omitted.

\begin{figure}[t]
\centering
\includegraphics[width=0.4\textwidth]{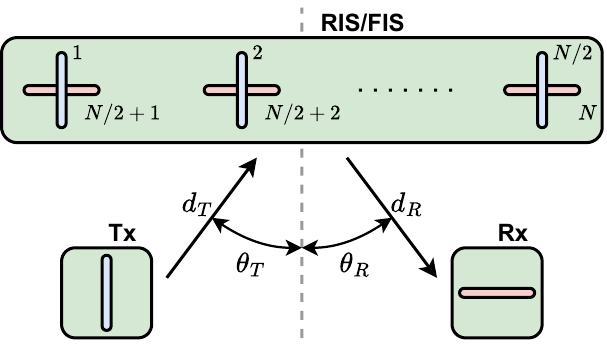}
\caption{Dual-polarized RIS/FIS-aided system where the transmitter and receiver have opposite polarization.}
\label{fig:system}
\end{figure}

The channels $\mathbf{h}_{R}$ and $\mathbf{h}_{T}$ are modeled by assuming that the surface is a dual-polarized \gls{ula} in \gls{los} with both the transmitter and receiver.
The $N$ elements of the surface are numbered such that the first $N/2$ have vertical polarization and the last $N/2$ have horizontal polarization, as shown in Fig.~\ref{fig:system}.
The effect of the element polarization is captured in $\mathbf{h}_{R}$ and $\mathbf{h}_{T}$ by modeling them as
\begin{equation}
\mathbf{h}_{R}=\mathbf{p}_{R}\otimes\mathbf{g}_{R},\;\mathbf{h}_{T}=\mathbf{p}_{T}\otimes\mathbf{g}_{T},\label{eq:hRhT}
\end{equation}
where $\otimes$ is the Kronecker product, $\mathbf{p}_{R}\in\mathbb{R}^{1\times 2}$ and $\mathbf{p}_{T}\in\mathbb{R}^{2\times 1}$ capture the power imbalance between the two polarizations, and $\mathbf{g}_{R}\in\mathbb{C}^{1\times N/2}$ and $\mathbf{g}_{T}\in\mathbb{C}^{N/2\times 1}$ are the uni-polarized \gls{los} fading channels \cite{oes08}.
The vectors $\mathbf{p}_{R}$ and $\mathbf{p}_{T}$ depend on the polarizations of the transmitter and receiver antennas.
For a vertically or horizontally polarized receiver, $\mathbf{p}_{R}=[1,\sqrt{\chi}]$ or $\mathbf{p}_{R}=[\sqrt{\chi},1]$, respectively, where $0\leq\chi\leq1$ is the inverse of the \gls{xpd}.
Likewise, for a vertically or horizontally polarized transmitter, $\mathbf{p}_{T}=[1,\sqrt{\chi}]^T$ or $\mathbf{p}_{T}=[\sqrt{\chi},1]^T$, respectively.

The \gls{los} fading channels $\mathbf{g}_{R}$ and $\mathbf{g}_{T}$ are expressed as follows.
We place the $N$ elements of the surface on the $x$-axis, centered in $x=0$, so that the $n$th element has $x$ coordinate $x_n=(n-(N+1)/2)d_A$, with $d_A$ being the spacing between the antenna elements.
Denoting as $d_{R}$ the distance between the center of the surface and the receiver, and as $\theta_{R}\in[-\pi/2,\pi/2]$ the angle of the receiver with respect to the surface normal, the distance between the $n$th element and the receiver is $d_{R,n}=d_{R}-x_n\sin(\theta_R)$ assuming far-field.
Hence, the entries of $\mathbf{g}_{R}$, given by $[\mathbf{g}_{R}]_n=e^{-j2\pi d_{R,n}/\lambda}$, are
\begin{equation}
\left[\mathbf{g}_{R}\right]_n=e^{-j\frac{2\pi}{\lambda}\left[d_{R}-\left(n-\frac{N+1}{2}\right)d_A\sin(\theta_R)\right]},\label{eq:gR}
\end{equation}
as a function of the wavelength $\lambda=c/f$, where $c$ the speed of light, and $f$ the frequency.
Likewise, denoting as $d_{T}$ the distance between the center of the surface and the transmitter, and as $\theta_T\in[-\pi/2,\pi/2]$ the angle of the transmitter, the entries of $\mathbf{g}_{T}$ are written as
\begin{equation}
\left[\mathbf{g}_{T}\right]_n=e^{-j\frac{2\pi}{\lambda}\left[d_{T}-\left(n-\frac{N+1}{2}\right)d_A\sin(\theta_T)\right]}.\label{eq:gT}
\end{equation}

\subsection{Problem Formulation}

In this communication system, our metric is the received power, written as
\begin{equation}
P_R=P_T\left\vert\mathbf{h}_{R}\boldsymbol{\Theta}\mathbf{h}_{T}-\mathbf{h}_{R}\mathbf{h}_{T}\right\vert^2,\label{eq:PR}
\end{equation}
where $P_T=\mathbb{E}[\vert x\vert^2]$ is the transmitted power, assumed $P_T=1$ for simplicity.
We compare two strategies for maximizing $P_R$.

\subsubsection{RIS}

With a \gls{ris}, the received power is maximized by reconfiguring $\boldsymbol{\Theta}$ depending on the channels $\mathbf{h}_{R}$ and $\mathbf{h}_{T}$, which are assumed to be known \cite{wu21}.
Formally, the problem is
\begin{equation}
\underset{\boldsymbol{\Theta}}{\mathsf{\mathrm{max}}}\;\;
P_R\;\;
\mathsf{\mathrm{s.t.}}\;\;
\mathbf{h}_{R}\text{ and }\mathbf{h}_{T}\text{ are fixed},
\end{equation}
where $\boldsymbol{\Theta}$ is reconfigurable subject to specific constraints.
For a \gls{d-ris}, we have $\boldsymbol{\Theta}=\text{diag}\left(e^{j\theta_1},\ldots,e^{j\theta_N}\right)$, where $\theta_n\in[0,2\pi)$ are the phase shifts that can be reconfigured.
For a \gls{bd-ris}, $\boldsymbol{\Theta}$ can be any unitary matrix, i.e., $\boldsymbol{\Theta}^H\boldsymbol{\Theta}=\mathbf{I}$, if only the lossless constraint is considered. 

\subsubsection{Movable Signals with FIS}

With movable signals, the received power is maximized by reconfiguring the frequency $f$ (or the wavelength $\lambda$), while the reflection matrix of the surface $\boldsymbol{\Theta}$ is fixed to a suitable predefined value.
Such a surface with a fixed reflection matrix is referred to as \gls{fis} \cite{ner25b}.
The optimization problem for movable signals is
\begin{equation}
\underset{\lambda}{\mathsf{\mathrm{max}}}\;\;
P_R\;\;
\mathsf{\mathrm{s.t.}}\;\;
\boldsymbol{\Theta}\text{ is fixed},\;\eqref{eq:hRhT},\;\eqref{eq:gR},\;\eqref{eq:gT},
\end{equation}
where $\boldsymbol{\Theta}$ is pre-optimized offline and fixed during deployment.
For a \gls{d-fis}, $\boldsymbol{\Theta}$ is set subject to $\boldsymbol{\Theta}=\text{diag}\left(e^{j\theta_1},\ldots,e^{j\theta_N}\right)$, while for a \gls{bd-fis}, it is subject to $\boldsymbol{\Theta}^H\boldsymbol{\Theta}=\mathbf{I}$.

In the two sections that follow, we compare the received power achievable through \gls{ris} and movable signals with \gls{fis}, by separately considering the two cases where the transmitter and receiver have the same and opposite polarization.

\section{Same Polarization at the Tx and Rx}

Assume the transmitter and receiver to have the same polarization, namely vertical, with no loss of generality.
Following \eqref{eq:hRhT}, this yields $\mathbf{h}_{R}=[\mathbf{g}_{R},\sqrt{\chi}\mathbf{g}_{R}]$ and $\mathbf{h}_{T}=[\mathbf{g}_{T}^T,\sqrt{\chi}\mathbf{g}_{T}^T]^T$.

\subsubsection{Diagonal RIS}

With a \gls{d-ris}, it is well known that the received power is maximized by setting $\boldsymbol{\Theta}$ as $\boldsymbol{\Theta}=\text{diag}(e^{j\theta_1},\ldots,e^{j\theta_N})$, with phase shifts $\theta_n=-\arg([\mathbf{h}_R]_n[\mathbf{h}_T]_n)+\arg(-\mathbf{h}_R\mathbf{h}_T)$, for $n=1,\dots,N$.
This global optimal solution gives
\begin{equation}
P_R^{\mathrm{DRIS}}=\left(\sum_{n=1}^N\left\vert\left[\mathbf{h}_{R}\right]_{n}\left[\mathbf{h}_{T}\right]_{n}\right\vert+\left\vert\mathbf{h}_{R}\mathbf{h}_{T}\right\vert\right)^2.\label{eq:PR-DRIS}
\end{equation}
Since $\mathbf{h}_{R}=[\mathbf{g}_{R},\sqrt{\chi}\mathbf{g}_{R}]$ and $\mathbf{h}_{T}=[\mathbf{g}_{T}^T,\sqrt{\chi}\mathbf{g}_{T}^T]^T$, we have $\vert[\mathbf{h}_{R}]_{n}[\mathbf{h}_{T}]_{n}\vert=1$, for $n=1,\ldots,N/2$, $\vert[\mathbf{h}_{R}]_{n}[\mathbf{h}_{T}]_{n}\vert=\chi$, for $n=N/2+1,\ldots,N$, and $\mathbf{h}_{R}\mathbf{h}_{T}=(1+\chi)\mathbf{g}_{R}\mathbf{g}_{T}$, giving
\begin{align}
P_R^{\mathrm{DRIS}}
&=\frac{\left(1+\chi\right)^2}{4}\left(N+2\left\vert\mathbf{g}_{R}\mathbf{g}_{T}\right\vert\right)^2.\label{eq:PR-same-DRIS}
\end{align}
Taking the expectation of \eqref{eq:PR-same-DRIS} over $\mathbf{g}_{R}$ and $\mathbf{g}_{T}$, whose entries are assumed to have independent
phases uniformly distributed in $[0,2\pi)$, the average received power $\bar{P}_R^{\mathrm{DRIS}}=\mathbb{E}[P_R^{\mathrm{DRIS}}]$ is
\begin{equation}
\bar{P}_R^{\mathrm{DRIS}}
=\frac{\left(1+\chi\right)^2}{4}\left(N^2+4N\mathbb{E}\left[\left\vert\mathbf{g}_{R}\mathbf{g}_{T}\right\vert\right]+4\mathbb{E}[\left\vert\mathbf{g}_{R}\mathbf{g}_{T}\right\vert^2]\right).
\end{equation}
Furthermore, since $\mathbf{g}_{R}\mathbf{g}_{T}$ is the sum of $N/2$ \gls{iid} random variables with mean 0 and variance 1, it can be approximated as a complex Gaussian with mean $0$ and variance $N/2$, i.e., $\mathbf{g}_{R}\mathbf{g}_{T}\sim\mathcal{CN}(0,N/2)$, by the Central Limit Theorem.
Hence, by exploiting $\mathbb{E}[\vert\mathbf{g}_{R}\mathbf{g}_{T}\vert]=\sqrt{\pi N/8}$ and $\mathbb{E}[\vert\mathbf{g}_{R}\mathbf{g}_{T}\vert^2]=N/2$, we obtain
\begin{equation}
\bar{P}_R^{\mathrm{DRIS}}
=\frac{\left(1+\chi\right)^2}{4}\left(N^2+\sqrt{2\pi N}N+2N\right),\label{eq:E-same-DRIS}
\end{equation}
showing that the received power with a \gls{d-ris} when transmitter and receiver have the same polarization scales with $\bar{P}_R^{\mathrm{DRIS}}\approx(1+\chi)^2N^2/4$.

\subsubsection{Beyond-Diagonal RIS}

With a \gls{bd-ris}, the achievable received power is given by
\begin{equation}
P_{R}^{\mathrm{BDRIS}}=\left(\left\Vert\mathbf{h}_{R}\right\Vert\left\Vert\mathbf{h}_{T}\right\Vert+\left\vert\mathbf{h}_{R}\mathbf{h}_{T}\right\vert\right)^{2},\label{eq:PR-BDRIS}
\end{equation}
which can be achieved by reconfiguring $\boldsymbol{\Theta}$ with the global optimal solution derived in \cite{ner24a}.
Since $\mathbf{h}_{R}=[\mathbf{g}_{R},\sqrt{\chi}\mathbf{g}_{R}]$ and $\mathbf{h}_{T}=[\mathbf{g}_{T}^T,\sqrt{\chi}\mathbf{g}_{T}^T]^T$, we have $\Vert\mathbf{h}_{R}\Vert^2=\Vert\mathbf{h}_{T}\Vert^2=(1+\chi)N/2$ and $\mathbf{h}_{R}\mathbf{h}_{T}=(1+\chi)\mathbf{g}_{R}\mathbf{g}_{T}$, giving that $P_R^{\mathrm{BDRIS}}$ is equal to the received power with \gls{d-ris} in \eqref{eq:PR-same-DRIS}.
Therefore, the average received power achievable with a \gls{bd-ris} $\bar{P}_R^{\mathrm{BDRIS}}=\mathbb{E}[P_R^{\mathrm{BDRIS}}]$ is the same as the average received power with a \gls{d-ris} given by \eqref{eq:E-same-DRIS}, scaling with $\bar{P}_R^{\mathrm{BDRIS}}\approx(1+\chi)^2N^2/4$.
The fact that \gls{d-ris} and \gls{bd-ris} achieve the same performance in dual-polarized systems when the transmitter and receiver have the same polarization was also derived in \cite{ner25a}, although the specular reflection was neglected in \cite{ner25a}.

\subsubsection{Diagonal FIS}
\label{sec:same-DFIS}

With movable signals and \gls{d-fis}, the reflection matrix of the \gls{fis} is pre-optimized offline subject to $\boldsymbol{\Theta}=\text{diag}\left(e^{j\theta_1},\ldots,e^{j\theta_N}\right)$, while the signal frequency $f$ is dynamically reconfigured.
We show how to fix $\boldsymbol{\Theta}$ and optimize $f$ by first deriving an upper bound on the received power, and then giving solutions for $\boldsymbol{\Theta}$ and $f$ which achieve it.
The received power with a \gls{d-fis} is upper bounded by
\begin{align}
P_R
&\leq\left(\sum_{n=1}^N\left\vert\left[\mathbf{h}_{R}\right]_{n}\left[\mathbf{h}_{T}\right]_{n}\right\vert+\left\vert\mathbf{h}_{R}\mathbf{h}_{T}\right\vert\right)^2\label{eq:up-same-DFIS2}\\
&=\left(1+\chi\right)^2\left(\frac{N}{2}+\left\vert\mathbf{g}_{R}\mathbf{g}_{T}\right\vert\right)^2
\leq\left(1+\chi\right)^2N^2,\label{eq:E-same-DFIS}
\end{align}
where \eqref{eq:up-same-DFIS2} is given by the triangle inequality, and \eqref{eq:E-same-DFIS} follows from $\vert[\mathbf{h}_{R}]_{n}[\mathbf{h}_{T}]_{n}\vert=1$, for $n=1,\ldots,N/2$, $\vert[\mathbf{h}_{R}]_{n}[\mathbf{h}_{T}]_{n}\vert=\chi$, for $n=N/2+1,\ldots,N$, $\mathbf{h}_{R}\mathbf{h}_{T}=(1+\chi)\mathbf{g}_{R}\mathbf{g}_{T}$, and $\vert\mathbf{g}_{R}\mathbf{g}_{T}\vert\leq N/2$.
To achieve the upper bound $P_R=(1+\chi)^2N^2$, the \gls{d-fis} matrix is fixed to $\boldsymbol{\Theta}^\star=-\mathbf{I}_{N}$, such that the term $\mathbf{h}_{R}\boldsymbol{\Theta}\mathbf{h}_{T}$ is aligned with the specular reflection $-\mathbf{h}_{R}\mathbf{h}_{T}$.
With $\boldsymbol{\Theta}=-\mathbf{I}_{N}$, the received power is
\begin{equation}
P_R
=\left\vert-\mathbf{h}_{R}\mathbf{h}_{T}-\mathbf{h}_{R}\mathbf{h}_{T}\right\vert^2
=4\left(1+\chi\right)^2\left\vert\mathbf{g}_{R}\mathbf{g}_{T}\right\vert^2,
\end{equation}
since $\mathbf{h}_{R}\mathbf{h}_{T}=(1+\chi)\mathbf{g}_{R}\mathbf{g}_{T}$, and therefore the signal frequency needs to be reconfigured to maximize $\vert\mathbf{g}_{R}\mathbf{g}_{T}\vert$, i.e., to ensure that $\vert\mathbf{g}_{R}\mathbf{g}_{T}\vert=N/2$.
Interestingly, this is achieved if the signal wavelength is reconfigured to $\lambda^\star=d_A\vert\sin(\theta_R)+\sin(\theta_T)\vert$, depending on the angles $\theta_R$ and $\theta_T$, with corresponding frequency
\begin{equation}
f^\star=\frac{c}{d_A\left\vert\sin(\theta_R)+\sin(\theta_T)\right\vert}.\label{eq:f}
\end{equation}
The optimality of this value can be verified by substituting $\lambda^\star$ into the expressions of $\mathbf{g}_{R}$ and $\mathbf{g}_{T}$ in \eqref{eq:gR} and \eqref{eq:gT} and verifying that it leads to $\vert\mathbf{g}_{R}\mathbf{g}_{T}\vert=N/2$.
Therefore, by reconfiguring the signal frequency as in \eqref{eq:f} a \gls{d-fis} with reflection matrix $\boldsymbol{\Theta}=-\mathbf{I}_{N}$ can achieve a received power $P_R^{\mathrm{DFIS}}=(1+\chi)^2N^2$.
Note that this value is four times higher than the received power achieved by a \gls{d-ris} $\bar{P}_R^{\mathrm{DRIS}}\approx(1+\chi)^2N^2/4$, and can be obtained without reconfiguring the \gls{em} properties of the surface.
Movable signals can achieve higher performance than \gls{ris} since they control both terms $\mathbf{h}_{R}\boldsymbol{\Theta}\mathbf{h}_{T}$ and $-\mathbf{h}_{R}\mathbf{h}_{T}$ in the received power expression, while \gls{ris} only controls $\mathbf{h}_{R}\boldsymbol{\Theta}\mathbf{h}_{T}$.

\subsubsection{Beyond-Diagonal FIS}

With movable signals and \gls{bd-fis}, the reflection matrix of the \gls{fis} is pre-optimized offline subject to $\boldsymbol{\Theta}^H\boldsymbol{\Theta}=\mathbf{I}$, while the signal frequency $f$ is dynamically reconfigured to maximize the received power.
In this case, the received power is upper bounded by
\begin{align}
P_R
&\leq\left(\left\Vert\mathbf{h}_{R}\right\Vert\left\Vert\mathbf{h}_{T}\right\Vert+\left\vert\mathbf{h}_{R}\mathbf{h}_{T}\right\vert\right)^2\label{eq:up-same-BDFIS2}\\
&=\left(1+\chi\right)^2\left(\frac{N}{2}+\left\vert\mathbf{g}_{R}\mathbf{g}_{T}\right\vert\right)^2
\leq\left(1+\chi\right)^2N^2,\label{eq:E-same-BDFIS}
\end{align}
where \eqref{eq:up-same-BDFIS2} is due to the triangle inequality and the sub-multiplicity of the $\ell_2$-norm, and \eqref{eq:E-same-BDFIS} holds because $\Vert\mathbf{h}_{R}\Vert^2=\Vert\mathbf{h}_{T}\Vert^2=(1+\chi)N/2$, $\mathbf{h}_{R}\mathbf{h}_{T}=(1+\chi)\mathbf{g}_{R}\mathbf{g}_{T}$, and $\vert\mathbf{g}_{R}\mathbf{g}_{T}\vert\leq N/2$.
Therefore, the maximum received power achievable by a \gls{bd-fis} is the same as the received power of a \gls{d-fis} given by \eqref{eq:E-same-DFIS}.
This implies that $\boldsymbol{\Theta}=-\mathbf{I}$ is global optimal also for a \gls{bd-fis}, and the signal frequency can still be reconfigured as in \eqref{eq:f} to achieve a received power of $P_R^{\mathrm{BDFIS}}\approx(1+\chi)^2N^2$.

We have studied a dual-polarized \gls{ris}/\gls{fis}-aided system where the transmitter and receiver have the same polarization, considering \gls{d-ris}, \gls{bd-ris}, and movable signals with \gls{d-fis} and \gls{bd-fis}.
The scaling laws of the received power in these four cases are summarized in Tab.~\ref{tab}, from which we make two observations.
\textit{First}, \gls{bd-ris} and \gls{bd-fis} do not improve the received power over \gls{d-ris} and \gls{d-fis}, respectively.
\textit{Second}, movable signals with \gls{fis} can obtain four times higher received power than fixed-frequency signals with \gls{ris}.

\section{Opposite Polarization at the Tx and Rx}

Assume now the transmitter and receiver to have opposite polarization, namely vertical and horizontal, respectively, with no loss of generality.
Following \eqref{eq:hRhT}, this yields $\mathbf{h}_{R}=[\sqrt{\chi}\mathbf{g}_{R},\mathbf{g}_{R}]$ and $\mathbf{h}_{T}=[\mathbf{g}_{T}^T,\sqrt{\chi}\mathbf{g}_{T}^T]^T$.

\subsubsection{Diagonal RIS}

With a \gls{d-ris}, the maximum received power is given by \eqref{eq:PR-DRIS}.
Since now the channels are $\mathbf{h}_{R}=[\sqrt{\chi}\mathbf{g}_{R},\mathbf{g}_{R}]$ and $\mathbf{h}_{T}=[\mathbf{g}_{T}^T,\sqrt{\chi}\mathbf{g}_{T}^T]^T$, we have $\vert[\mathbf{h}_{R}]_{n}[\mathbf{h}_{T}]_{n}\vert=\sqrt{\chi}$, for $n=1,\ldots,N$, and $\mathbf{h}_{R}\mathbf{h}_{T}=2\sqrt{\chi}\mathbf{g}_{R}\mathbf{g}_{T}$, allowing us to simplify \eqref{eq:PR-DRIS} as
\begin{align}
P_R^{\mathrm{DRIS}}
&=\chi\left(N+2\left\vert\mathbf{g}_{R}\mathbf{g}_{T}\right\vert\right)^2.\label{eq:PR-oppo-DRIS}
\end{align}
Taking the expectation of \eqref{eq:PR-oppo-DRIS} over all possible channels $\mathbf{g}_{R}$ and $\mathbf{g}_{T}$, the average received power $\bar{P}_R^{\mathrm{DRIS}}=\mathbb{E}[P_R^{\mathrm{DRIS}}]$ is
\begin{equation}
\bar{P}_R^{\mathrm{DRIS}}
=\chi\left(N^2+4N\mathbb{E}\left[\left\vert\mathbf{g}_{R}\mathbf{g}_{T}\right\vert\right]+4\mathbb{E}[\left\vert\mathbf{g}_{R}\mathbf{g}_{T}\right\vert^2]\right),
\end{equation}
where we can make the approximation $\mathbf{g}_{R}\mathbf{g}_{T}\sim\mathcal{CN}(0,N/2)$, following the Central Limit Theorem.
Therefore, by using $\mathbb{E}[\vert\mathbf{g}_{R}\mathbf{g}_{T}\vert]=\sqrt{\pi N/8}$ and $\mathbb{E}[\vert\mathbf{g}_{R}\mathbf{g}_{T}\vert^2]=N/2$, we get
\begin{equation}
\bar{P}_R^{\mathrm{DRIS}}
=\chi\left(N^2+\sqrt{2\pi N}N+2N\right),\label{eq:E-oppo-DRIS}
\end{equation}
giving the scaling law of the received power in the presence of a \gls{d-ris} as $\bar{P}_R^{\mathrm{DRIS}}\approx\chi N^2$.

\begin{table}[t]
\centering
\caption{Scaling laws of the received power in\\dual-polarized RIS/FIS-aided systems.}
\begin{tabular}{|c|c|}
\hline
Same polariz. at Tx and Rx & Opposite polariz. at Tx and Rx\\
\hline
$\begin{aligned}
\bar{P}_R^{\mathrm{DRIS}}&=\left(1+\chi\right)^2N^2/4\\
\bar{P}_R^{\mathrm{BDRIS}}&=\left(1+\chi\right)^2N^2/4\\
P_R^{\mathrm{DFIS}}&=\left(1+\chi\right)^2N^2\\
P_R^{\mathrm{BDFIS}}&=\left(1+\chi\right)^2N^2
\end{aligned}$&
$\begin{aligned}
\bar{P}_R^{\mathrm{DRIS}}&=\chi N^2\\
\bar{P}_R^{\mathrm{BDRIS}}&=\left(1+\chi\right)^2N^2/4\\
P_R^{\mathrm{DFIS}}&=4\chi N^2\\
P_R^{\mathrm{BDFIS}}&=\left(1+\chi+2\sqrt{\chi}\right)^2N^2/4
\end{aligned}$\\
\hline
\end{tabular}
\label{tab}
\end{table}

\subsubsection{Beyond-Diagonal RIS}

In the presence of a \gls{bd-ris}, the maximum received power is given by \eqref{eq:PR-BDRIS}.
Since $\mathbf{h}_{R}=[\sqrt{\chi}\mathbf{g}_{R},\mathbf{g}_{R}]$ and $\mathbf{h}_{T}=[\mathbf{g}_{T}^T,\sqrt{\chi}\mathbf{g}_{T}^T]^T$ give $\Vert\mathbf{h}_{R}\Vert^2=\Vert\mathbf{h}_{T}\Vert^2=(1+\chi)N/2$ and $\mathbf{h}_{R}\mathbf{h}_{T}=2\sqrt{\chi}\mathbf{g}_{R}\mathbf{g}_{T}$, \eqref{eq:PR-BDRIS} becomes
\begin{equation}
P_R^{\mathrm{BDRIS}}
=\left(\left(1+\chi\right)\frac{N}{2}+2\sqrt{\chi}\left\vert\mathbf{g}_{R}\mathbf{g}_{T}\right\vert\right)^2.\label{eq:PR-oppo-BDRIS}
\end{equation}
Taking the expectation of \eqref{eq:PR-oppo-BDRIS} over all possible channels $\mathbf{g}_{R}$ and $\mathbf{g}_{T}$, the average received power $\bar{P}_R^{\mathrm{BDRIS}}=\mathbb{E}[P_R^{\mathrm{BDRIS}}]$ is
\begin{multline}
\bar{P}_R^{\mathrm{BDRIS}}
=\frac{\left(1+\chi\right)^2}{4}N^2+2\left(1+\chi\right)\sqrt{\chi}N\mathbb{E}\left[\left\vert\mathbf{g}_{R}\mathbf{g}_{T}\right\vert\right]\\
+4\chi\mathbb{E}[\left\vert\mathbf{g}_{R}\mathbf{g}_{T}\right\vert^2].
\end{multline}
Furthermore, by approximating $\mathbf{g}_{R}\mathbf{g}_{T}$ as $\mathbf{g}_{R}\mathbf{g}_{T}\sim\mathcal{CN}(0,N/2)$ following the Central Limit Theorem, we can apply $\mathbb{E}[\vert\mathbf{g}_{R}\mathbf{g}_{T}\vert]=\sqrt{\pi N/8}$ and $\mathbb{E}[\vert\mathbf{g}_{R}\mathbf{g}_{T}\vert^2]=N/2$ to obtain
\begin{equation}
\bar{P}_R^{\mathrm{BDRIS}}
=\frac{\left(1+\chi\right)^2}{4}N^2+\left(1+\chi\right)\sqrt{\frac{\pi}{2}\chi N}N+2\chi N,\label{eq:E-oppo-BDRIS}
\end{equation}
showing that the received power with a \gls{bd-ris} scales with $\bar{P}_R^{\mathrm{BDRIS}}\approx(1+\chi)^2N^2/4$, always higher than $\bar{P}_R^{\mathrm{DRIS}}\approx\chi N^2$ when $\chi\neq1$.

\subsubsection{Diagonal FIS}

When considering movable signals and \gls{d-fis}, the received power is upper bounded by
\begin{align}
P_R
&\leq\left(\sum_{n=1}^N\left\vert\left[\mathbf{h}_{R}\right]_{n}\left[\mathbf{h}_{T}\right]_{n}\right\vert+\left\vert\mathbf{h}_{R}\mathbf{h}_{T}\right\vert\right)^2\label{eq:up-oppo-DFIS2}\\
&=\left(\sqrt{\chi}N+2\sqrt{\chi}\vert\mathbf{g}_{R}\mathbf{g}_{T}\vert\right)^2
\leq4\chi N^2,\label{eq:E-oppo-DFIS}
\end{align}
where \eqref{eq:up-oppo-DFIS2} is given by the triangle inequality, and \eqref{eq:E-oppo-DFIS} follows from $\vert[\mathbf{h}_{R}]_{n}[\mathbf{h}_{T}]_{n}\vert=1$, for $n=1,\ldots,N/2$, $\vert[\mathbf{h}_{R}]_{n}[\mathbf{h}_{T}]_{n}\vert=\chi$, for $n=N/2+1,\ldots,N$, $\mathbf{h}_{R}\mathbf{h}_{T}=2\sqrt{\chi}\mathbf{g}_{R}\mathbf{g}_{T}$, and $\vert\mathbf{g}_{R}\mathbf{g}_{T}\vert\leq N/2$.
As discussed in Section~\ref{sec:same-DFIS}, the upper bound $P_R=4\chi N^2$ can be achieved by fixing the \gls{d-fis} matrix to $\boldsymbol{\Theta}^\star=-\mathbf{I}_{N}$.
With $\boldsymbol{\Theta}=-\mathbf{I}_{N}$, the received power is
\begin{equation}
P_R
=\left\vert-\mathbf{h}_{R}\mathbf{h}_{T}-\mathbf{h}_{R}\mathbf{h}_{T}\right\vert^2
=16\chi\left\vert\mathbf{g}_{R}\mathbf{g}_{T}\right\vert^2,
\end{equation}
since $\mathbf{h}_{R}\mathbf{h}_{T}=2\sqrt{\chi}\mathbf{g}_{R}\mathbf{g}_{T}$, and therefore the signal frequency needs to be optimized to maximize $\vert\mathbf{g}_{R}\mathbf{g}_{T}\vert$.
As discussed in Section~\ref{sec:same-DFIS}, $\vert\mathbf{g}_{R}\mathbf{g}_{T}\vert$ is maximized when the signal frequency is set as in \eqref{eq:f}, depending on the angles $\theta_R$ and $\theta_T$.
Therefore, by using a signal frequency as in \eqref{eq:f} a \gls{d-fis} with reflection matrix $\boldsymbol{\Theta}=-\mathbf{I}_{N}$ can achieve a received power $P_R^{\mathrm{DFIS}}=4\chi N^2$.
Note that this value is four times higher than the received power achieved by a \gls{d-ris} $\bar{P}_R^{\mathrm{DRIS}}\approx\chi N^2$, while requiring no reconfiguration of \gls{em} properties of the surface.

\begin{figure}[t]
\centering
\includegraphics[width=0.38\textwidth]{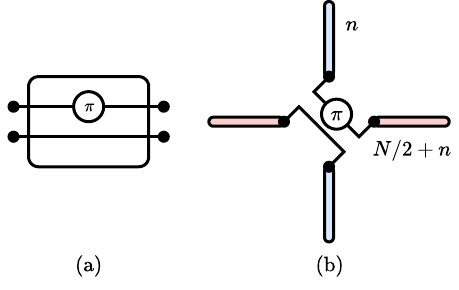}
\caption{(a) A phase shifter with phase $\pi$ as an example of a $2$-port network with scattering matrix $[[0,-1]^T,[-1,0]^T]$, and (b) implementation of the $n$th group of the BD-FIS in Section~\ref{sec:oppo-BDFIS}.}
\label{fig:schematic}
\end{figure}

\subsubsection{Beyond-Diagonal FIS}
\label{sec:oppo-BDFIS}

With movable signals and \gls{bd-fis}, the received power is upper bounded by
\begin{align}
P_R
&\leq\left(\left\Vert\mathbf{h}_{R}\right\Vert\left\Vert\mathbf{h}_{T}\right\Vert+\left\vert\mathbf{h}_{R}\mathbf{h}_{T}\right\vert\right)^2\label{eq:up-oppo-BDFIS1}\\
&=\left(\left(1+\chi\right)\frac{N}{2}+2\sqrt{\chi}\left\vert\mathbf{g}_{R}\mathbf{g}_{T}\right\vert\right)^2
\leq\frac{\left(1+\chi+2\sqrt{\chi}\right)^2}{4}N^2,\label{eq:E-oppo-BDFIS}
\end{align}
where \eqref{eq:up-oppo-BDFIS1} is due to the triangle inequality and the sub-multiplicity of the $\ell_2$-norm, and \eqref{eq:E-oppo-BDFIS} holds since $\Vert\mathbf{h}_{R}\Vert^2=\Vert\mathbf{h}_{T}\Vert^2=(1+\chi)N/2$, $\mathbf{h}_{R}\mathbf{h}_{T}=2\sqrt{\chi}\mathbf{g}_{R}\mathbf{g}_{T}$, and $\vert\mathbf{g}_{R}\mathbf{g}_{T}\vert\leq N/2$.
We now show how to fix the \gls{bd-fis} matrix and optimize the signal frequency to achieve this upper bound.
The \gls{bd-fis} matrix is fixed to
\begin{equation}
\boldsymbol{\Theta}=
\begin{bmatrix}
\mathbf{0}_{N/2} & -\mathbf{I}_{N/2}\\
-\mathbf{I}_{N/2} & \mathbf{0}_{N/2}
\end{bmatrix},\label{eq:T}
\end{equation}
which means we are considering a group-connected architecture with $N/2$ groups, each including 2 elements \cite{she22} (note that \eqref{eq:T} is a block diagonal matrix with $2\times2$ blocks being $[[0,-1]^T,[-1,0]^T]$ if its columns are permuted).
The groups are formed by two co-located elements having opposite polarization and all have a $2\times2$ scattering matrix given by $[[0,-1]^T,[-1,0]^T]$.
Since an example of $2$-port network having scattering matrix $[[0,-1]^T,[-1,0]^T]$ is a phase shifter with phase $\pi$ (Fig.~\ref{fig:schematic}(a)), such a \gls{bd-fis} can be implemented by connecting each pair of co-located elements via a phase shifter with phase $\pi$ (Fig.~\ref{fig:schematic}(b)).
In this way, the signal incident on the vertically polarized elements is reflected from the horizontally polarized ones, and vice versa, and the \gls{bd-fis} is a polarization converter.
With \eqref{eq:T}, the received power is
\begin{align}
P_R
&=\left\vert\mathbf{h}_{R}
\begin{bmatrix}
\mathbf{0}_{N/2} & -\mathbf{I}_{N/2}\\
-\mathbf{I}_{N/2} & \mathbf{0}_{N/2}
\end{bmatrix}
\mathbf{h}_{T}-\mathbf{h}_{R}\mathbf{h}_{T}\right\vert^2\\
&=\left(1+\chi+2\sqrt{\chi}\right)^2\left\vert\mathbf{g}_{R}\mathbf{g}_{T}\right\vert^2,
\end{align}
where we used $\mathbf{h}_{R}=[\sqrt{\chi}\mathbf{g}_{R},\mathbf{g}_{R}]$ and $\mathbf{h}_{T}=[\mathbf{g}_{T}^T,\sqrt{\chi}\mathbf{g}_{T}^T]^T$.
The signal frequency needs now to be reconfigured to maximize $\vert\mathbf{g}_{R}\mathbf{g}_{T}\vert$, i.e., to ensure that $\vert\mathbf{g}_{R}\mathbf{g}_{T}\vert=N/2$, which can be achieved by setting it as in \eqref{eq:f}.
Therefore, by reconfiguring the signal frequency as in \eqref{eq:f} a \gls{bd-fis} with reflection matrix given by \eqref{eq:T} achieves a received power $P_R^{\mathrm{BDFIS}}=(1+\chi+2\sqrt{\chi})^2N^2/4$.
Remarkably, this value is higher than the received power achieved by a \gls{d-fis} $P_R^{\mathrm{DFIS}}=4\chi N^2$ when $\chi\neq1$, and always higher than the received power of \gls{bd-ris}.

We have studied a dual-polarized \gls{ris}/\gls{fis}-aided system where the transmitter and receiver have opposite polarization.
The scaling laws of the received power achieved when considering \gls{d-ris}, \gls{bd-ris}, and movable signals with \gls{d-fis} and \gls{bd-fis} are summarized in Tab.~\ref{tab}.
We observe that \gls{bd-ris} and \gls{bd-fis} improve the received power over \gls{d-ris} and \gls{d-fis} when $\chi\neq1$, respectively, and that movable signals with \gls{fis} always outperform fixed-frequency signals with \gls{ris}.

\begin{figure}[t]
\centering
\includegraphics[width=0.38\textwidth]{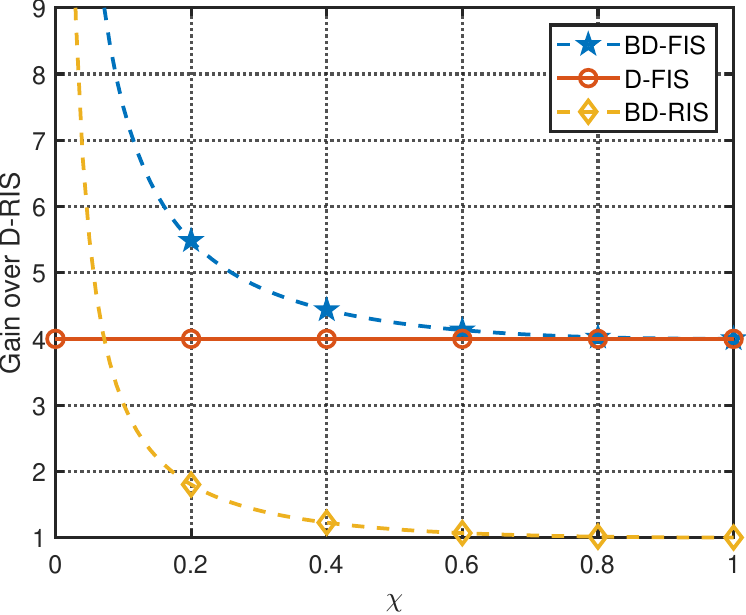}
\caption{Gain of BD-RIS, D-FIS, and BD-RIS over D-RIS as a function of $\chi$, when the transmitter and receiver have opposite polarization.}
\label{fig:gain}
\end{figure}

\section{Gain over Diagonal RIS}

In this section, we quantify the gain of \gls{bd-ris}, \gls{d-fis}, and \gls{bd-fis} over \gls{d-ris}, defined as $G^{\mathrm{X}}=\lim_{N\to+\infty}\bar{P}_R^{\mathrm{X}}/\bar{P}_R^{\mathrm{DRIS}}$, where $\mathrm{X}\in\{\mathrm{BDRIS},\;\mathrm{DFIS},\;\mathrm{BDFIS}\}$.
When the transmitter and receiver have the same polarization, these gains are given by $G^{\mathrm{BDRIS}}=1$, $G^{\mathrm{DFIS}}=4$, and $G^{\mathrm{BDFIS}}=4$, which be shown by using the scaling laws in \eqref{eq:E-same-DRIS}, \eqref{eq:E-same-DFIS}, and \eqref{eq:E-same-BDFIS}.
Besides, when the transmitter and receiver have opposite polarization, these gains are $G^{\mathrm{BDRIS}}=(1+\chi)^2/(4\chi)$, $G^{\mathrm{DFIS}}=4$, and $G^{\mathrm{BDFIS}}=(1+\chi+2\sqrt{\chi})^2/(4\chi)$, which are obtained by using the scaling laws in \eqref{eq:E-oppo-DRIS}, \eqref{eq:E-oppo-BDRIS}, \eqref{eq:E-oppo-DFIS}, and \eqref{eq:E-oppo-BDFIS}.

Focusing on systems where the transmitter and receiver have opposite polarization, we report the gains $G^{\mathrm{BDRIS}}$, $G^{\mathrm{DFIS}}$, and $G^{\mathrm{BDFIS}}$ as a function of $\chi$ in Fig.~\ref{fig:gain}.
We make the following three observations.
\textit{First}, the gain of \gls{bd-ris} over \gls{d-ris} tends to infinity when $\chi=0$, decreases with $\chi$, and is $1$ when $\chi=1$.
\Gls{bd-ris} is therefore particularly beneficial for small values of $\chi$, as also explained in \cite{ner25a}.
\textit{Second}, the gain of \gls{d-fis} over \gls{d-ris} is always $4$ and does not vary with $\chi$, as for uni-polarized systems \cite{ner25b}.
\textit{Third}, the gain of \gls{bd-fis} over \gls{d-ris} tends to infinity when $\chi=0$, decreases with $\chi$, and reaches $4$ when $\chi=1$.
Therefore, \gls{bd-fis} with movable signals is an appealing strategy for achieving substantially higher performance than \gls{ris} in dual-polarized systems.
Since its reconfigurability lies in the signal frequency, \gls{bd-fis} has a fixed reflection matrix, which is easy to implement and does not require control hardware.

\section{Conclusion}

We have investigated the performance limits of \gls{siso} systems aided by dual-polarized intelligent surfaces.
Specifically, we have compared two strategies: \gls{ris}, where the reflection properties of the surface are reconfigured, and movable signals with \gls{fis}, where the signal frequency is reconfigured while the surface has fixed reflection properties.
We show that movable signals with \gls{fis} always outperform \gls{ris}, by achieving at least a four times higher received power.
When the transmitter and receiver have opposite polarization, a \gls{fis} with a reflection matrix not limited to be diagonal, namely a \gls{bd-fis}, remarkably further increases the performance.
Future research could explore \gls{bd-fis} architectures that are optimal for any polarization of the transmitter and receiver, and the optimization of the frequency of movable signals under realistic spectrum constraints and channel models, e.g., in the near-field.

\bibliographystyle{IEEEtran}
\bibliography{IEEEabrv,main}

\end{document}